\documentclass[twocolumn,showpacs,showkeys,amsmath,amssymb,prl]{revtex4}

\usepackage{graphics}
\usepackage{bm}

\begin{document}

\title{Decoherence of orbital angular momentum entanglement in a turbulent atmosphere}

\author{Filippus S. Roux}

\affiliation{CSIR National Laser Centre, PO Box 395, Pretoria 0001, South Africa} 

\email{fsroux@csir.co.za} 

\begin{abstract}
The evolution of an entangled photon state propagating through a turbulent atmosphere is formulated in terms of a set of coupled first order differential equations, by using an infinitesimal propagation approach. The orbital angular momentum (OAM) basis is used to described the density matrix of the state. Although the analysis is done in the paraxial limit for a monochromatic optical field, the formalism is comprehensive in the sense that it does not require any assumptions about the strength of the turbulence and it can incorporate any spectral model for the turbulence. As a comparative example the case of entangled qubit OAM biphoton states is considered.
\end{abstract}



\maketitle

Orbital angular momentum (OAM) as a basis for entangled photon states recently became popular due to the potential for the higher dimensional quantum information processing and communication that it enables \cite{zeil1,qkdn,torres}. Since single mode optical fibre (the workhorse in the optical fibre communication infrastructure) only supports modes with zero OAM, free-space is the natural choice for OAM based quantum communication. Unfortunately OAM entanglement suffers decoherence in a turbulent atmosphere --- the random phase modulations, induced by the spatial fluctuation in the refractive index, scatter the initial OAM states into other OAM states.

Some studies have been done on the effect of scintillation on classical optical beams with specific OAM modes
\cite{turboam0,turboam1,turboam2,turboam3,turboam4}. There has also been some work done on quantum OAM states propagating through turbulence \cite{qturb1,qturb3}, however, these studies modeled the turbulence by a single phase-only transmission function that quantifies the turbulence by a single parameter, the Fried parameter $r_0$. Reliable calculations of the effects of turbulence on the propagation of light require more accurate models, which are usually represented by the power spectral densities of the refractive index fluctuation \cite{Hill,Frehlich}.

The approach that is followed in this paper, is to consider the incremental change in the density operator during an infinitesimal propagation through the turbulent atmosphere and to express the result as a differential equation. The fluctuation in the refractive index is extremely small compared to the average refractive index. As a result the distance over which the effect of the medium becomes visible is much longer than the correlation distance of the medium. One can therefore ignore correlations in the medium along the propagation distance, resulting in a Markov approximation. This approach is reminiscent of a Master equation approach, but instead of a time derivative, we obtain a derivative with respect to propagation distance. The analysis is restricted to monochromatic optical fields under the paraxial approximation.

When propagating through a turbulent atmosphere, the density operator for any photon state expressed in terms of an OAM basis, for example for a single photon given by
\begin{equation}
\rho(z) = \sum_{mn} |m \rangle\ \rho_{mn}(z)\ \langle n | ,
\label{rho}
\end{equation}
would change due to the distortion of the OAM modes. Here $\rho_{mn}(z)$ is the density matrix elements and $|m\rangle$, etc.\ denotes the OAM basis elements. (The two indices of the OAM modes --- radial index $p$ and azimuthal index $l$ --- are for the sake of notational simplicity combined into one index.) Since the turbulence model is represented by a spectral function, it is necessary to express the OAM basis elements as momentum-space wave functions $G_m({\bf K},z) = \langle{\bf K}|m,z\rangle$, where ${\bf K}$ is the two-dimensional spatial frequency vector on the transverse momentum-space. Formally the OAM wave functions $G_m({\bf K},z)$ are given by two-dimensional Fourier transforms of the Laguerre-Gaussian modal functions. The explicit $z$-dependence is a reminder that the OAM modes are functions of the propagation distance. For any given $z$, the OAM modes $|m,z\rangle$ form a complete two-dimensional orthogonal basis, and, by implication, so do their momentum-space wave functions $G_m({\bf K},z)$.

Another consequence of the extremely small index fluctuation is that the wave equation for monochromatic light propagating through turbulence separates into two terms in the paraxial limit:\ one for free-space propagation without turbulence and another for the phase modulation induced by the index fluctuation. For propagation over an infinitesimal distance $dz$, the momentum-space wave function is transformed as follows
\begin{eqnarray}
G_m({\bf K},z) & \stackrel{dz}{\longrightarrow} & G_m({\bf K},z) + \frac{i dz}{2k} \left[ |{\bf K}|^2 G_m({\bf K},z) \right. \nonumber \\ & & \left. - 2k^2 N({\bf K},z) \star G_m({\bf K},z) \right] ,
\label{hsedz}
\end{eqnarray}
where $\star$ represents convolution, and $k$ is the wave number ($k\neq|{\bf k}|$). The two-dimensional Fourier transform of the refractive index fluctuation is defined as
\begin{equation}
N({\bf K},z) = \int_{-\infty}^{\infty} \tilde{\chi}({\bf k}) \left[ \frac{\Phi_0({\bf k})}{\Delta_k^3} \right]^{1/2} \exp(-ik_z z)\ \frac{{\rm d}k_z}{2\pi} ,
\label{nspek}
\end{equation}
where ${\bf k}$ is the full three-dimensional spatial frequency vector, $\Phi_0({\bf k})$ is the three-dimensional power spectral density of the refractive index fluctuation, and $\tilde{\chi}({\bf k})$ is a three-dimensional random complex spectral function, with $\Delta_k$ being its coherence width in the frequency domain. Since the index fluctuation is a real-valued function, $\tilde{\chi}^*({\bf k})=\tilde{\chi}(-{\bf k})$. Furthermore,
\begin{equation}
\left\langle \tilde{\chi}({\bf k}_1) \tilde{\chi}^*({\bf k}_2) \right\rangle = \left( 2\pi\Delta_k \right)^3\ \delta_3({\bf k}_1-{\bf k}_2) ,
\label{verwagt}
\end{equation}
where $\delta_3({\bf k})$ denotes a Dirac-delta function in three dimensions.

It is important to note that, while $G_m({\bf K},z)$ represents an orthogonal basis, the distortions caused by the scintillation process denoted in Eq.~(\ref{hsedz}), produce a momentum-space wave function that is not an element of the orthogonal basis anymore. {\it The point of this approach is to re-express the resulting distorted wave function in terms of the orthogonal OAM basis and to incorporate the coefficients of this expansion into the density matrix elements.}

First we consider the case for a single photon. The resulting expressions can then be used to obtain equivalent expressions for entangled multi-photon states. The density matrix for a single photon, in terms of the momentum-space wave functions, is given by
\begin{eqnarray}
\rho(z) & = & \sum_{mn} \int_{-\infty}^{\infty} |{\bf K}_1 \rangle\ G_m({\bf K}_1) \rho_{mn}(z) \nonumber \\ & & \times G_n^*({\bf K}_2)\ \langle {\bf K}_2|\ \frac{{\rm d}^2K_1}{4\pi^2} \frac{{\rm d}^2K_2}{4\pi^2} ,
\label{rhog}
\end{eqnarray}
where we now dropped the $z$-dependence from the momentum-space wave functions.

Applying the propagation process of Eq.~(\ref{hsedz}) to the density operator in Eq.~(\ref{rhog}) repeatedly and averaging over the index fluctuation using Eqs.~(\ref{nspek}) and (\ref{verwagt}), one obtains an expression for the change in the density operator,
\begin{widetext}
\begin{eqnarray}
\partial_z \rho(z) & = & \sum_{mnpq} |m \rangle \rho_{pq}(z) \langle n |\ \int_{-\infty}^{\infty} G_m^*({\bf K}_1) G_n({\bf K}_2) \left[ k^2 \int_{-\infty}^{\infty}\!\! \Phi_1({\bf K}') G_p({\bf K}_1-{\bf K}') G_q^*({\bf K}_2-{\bf K}')\ \frac{{\rm d}^2K'}{4\pi^2} \right. \nonumber \\ & & \left. - k^2 G_p({\bf K}_1) G_q^*({\bf K}_2) \int_{-\infty}^{\infty}\!\! \Phi_1({\bf K}')\ \frac{{\rm d}^2K'}{4\pi^2} +\frac{i}{2k} \left( |{\bf K}_1|^2-|{\bf K}_2|^2 \right) G_p({\bf K}_1) G_q^*({\bf K}_2) \right]\ \frac{{\rm d}^2K_1}{4\pi^2} \frac{{\rm d}^2K_2}{4\pi^2} ,
\label{binfprop}
\end{eqnarray}
\end{widetext}
where $\Phi_1({\bf K})=\Phi_0({\bf K},k_z=0)$, which follows from the Markov approximation.

The first term in the square brackets in Eq.~(\ref{binfprop}) requires the evaluation of integrals of the form
\begin{equation}
W_{mn}({\bf K},z) = \int_{-\infty}^{\infty} G_m({\bf K}_1) G_n^*({\bf K}_1-{\bf K})\ \frac{{\rm d}^2K_1}{4\pi^2} .
\label{wees}
\end{equation}
The $W_{mn}$'s are then used in the remaining integrals of the first term to give
\begin{equation}
L_{mnpq} = k^2\!\!\int_{-\infty}^{\infty}\!\!\! \Phi_1({\bf K}) W_{mp}^*({\bf K}) W_{nq}({\bf K})\ \frac{{\rm d}^2K}{4\pi^2} .
\label{intg4}
\end{equation}

The integrals over ${\bf K}_1$ and ${\bf K}_2$ for the second term in the square brackets in Eq.~(\ref{binfprop}) represent pure orthogonality conditions and give rise to Kronecker delta functions. So the second term becomes $-\delta_{m,p}\delta_{n,q} L_T$, where
\begin{equation}
L_T = k^2 \int_{-\infty}^{\infty}\!\! \Phi_1({\bf K})\ \frac{{\rm d}^2K}{4\pi^2} .
\label{haa0}
\end{equation}

The third term in the square brackets in Eq.~(\ref{binfprop}), which is associated with the free-space propagation of OAM modes, is given by
\begin{eqnarray}
V_{mnpq} & = & \frac{i}{2k} \int_{-\infty}^{\infty}\!\!\! \left( |{\bf K}_1|^2-|{\bf K}_2|^2 \right) G_p({\bf K}_1) G_m^*({\bf K}_1)
\nonumber \\ & & \times G_n({\bf K}_2) G_q^*({\bf K}_2)\ \frac{{\rm d}^2K_1}{4\pi^2} \frac{{\rm d}^2K_2}{4\pi^2} .
\label{vees}
\end{eqnarray}
It is an orthogonality condition with respect to the azimuthal indices of the OAM modes, but not with respect to the radial indices. The latter is a result of the fact that the OAM modes depend on the propagation distance.

Considering the case of an entangled biphoton, one still finds integrals of the forms given in Eqs.~(\ref{wees}-\ref{vees}). The resulting expression for the density matrix elements in the case where one of the two photons propagates through turbulence, while the other propagates through free-space without turbulence, is given by
\begin{eqnarray}
\partial_z \rho_{mnpq} & = & V_{mnrs} \rho_{rspq} + V_{pqrs} \rho_{mnrs} \nonumber \\ & & + L_{mnrs} \rho_{rspq} - L_T \rho_{mnpq} ,
\label{bmaster}
\end{eqnarray}
where repeated indices are summed over. The equation in Eq.~(\ref{bmaster}) has the form of a Lindblad equation where the first two terms on the right-hand side represent the Hamiltonian term and the last two terms represent the dissipative terms.

In general Eq.~(\ref{bmaster}) represents an infinite set of coupled first order differential equations. Even if the initial state contains only a few lower order modes, the turbulence will cause these modes to be coupled into all other modes. Subsequently the other modes will couple back into the original modes. Truncating the set of equations, one inevitably excludes part of the coupling among all the different modes. However, this coupling should become progressively smaller for higher order modes. Hence, one may be able to truncate the set at some point while retaining the dominant inter-modal coupling.

To compare this result with previous work \cite{qturb1}, we consider a severely truncated case, only retaining modes of the lowest radial index ($p=0$) and with azimuthal indices of the same magnitude $l=\pm q$. We'll consider three cases where $q=1,2,3$, respectively. The truncation implies that the trace of the density matrix is not equal to 1 anymore. One can perform a normalization on the truncated density matrix to ensure that its trace remains 1. However, the reduced trace gives an indication of the loss of information to the higher order modes.

The formalism allows one to include any spectral model $\Phi_0({\bf k})$ for the turbulence, however, for the sake of comparison we neglect the effect of the inner and outer scales. Here we use the von Karman spectrum \cite{scintbook},
\begin{equation}
\Phi_0({\bf k}) = \frac{0.033 C_n^2}{(|{\bf k}|^2+\kappa_0^2)^{11/6}} ,
\label{karman}
\end{equation}
where $C_n^2$ is the structure constant for the turbulence and $\kappa_0$ is inversely proportional to the outer scale of the turbulence. The outer scale will help us to regularize the integrals, but will disappear from the final expressions. Substituting Eq.~(\ref{karman}) into Eq.~(\ref{haa0}), one obtains,
\begin{equation}
L_T = 0.1244 C_n^2 \kappa_0^{-5/3} + {\rm O}(1) .
\label{haa0a}
\end{equation}

For the case under consideration $V_{mnpq}(z)=0$. After evaluating the integrals for $L_{mnrs}$ one finds that, in the limit of large outer scale, the only nonzero terms are
\begin{eqnarray}
L_{q,q,q,q}(z) & = & L_{q,-q,q,-q}(z) = L_{-q,q,-q,q}(z) \nonumber \\ & = & L_{-q,-q,-q,-q}(z) = L_T - A_q h(z)
\label{haa1}
\end{eqnarray}
and
\begin{equation}
L_{q,q,-q,-q}(z) = L_{-q,-q,q,q}(z) = B_q h(z) ,
\label{haa2}
\end{equation}
where $A_q$ and $B_q$ are positive constants that only depend on $q$ (see Table~\ref{indkon}), and $h(z)$ is the same function for all the terms. It contains all the dimension parameters
\begin{equation}
h(z) = \frac{1}{z_R} \left(C_n^2 \omega_0^{2/3} \right) \left(\frac{\lambda}{\pi\omega_0}\right)^{-3} \left(1+\frac{z^2}{z_R^2}\right)^{5/6} ,
\label{haaz}
\end{equation}
where $z_R$ is the Rayleigh range ($\pi\omega_0^2/\lambda$), $\omega_0$ is the radius of the beam waist and $\lambda$ is the wavelength.

From Eq.~(\ref{haa0a}) one can see that $L_T$ diverge in the limit of large outer scale. However, the nonzero elements for the $L_T$-term in Eq.~(\ref{bmaster}) are the same as those in Eq.~(\ref{haa1}). As a result all the $L_T$-terms cancel exactly and the outer scale drops out of the final expression.

Provided that the turbulence is not too weak and that the beam waist is not too small, the concurrence decays to zero over a distance much shorter than the Rayleigh range, which allows one to assume that $1+z^2/z_R^2\approx 1$. Under these circumstances one can express the integral of $h(z)$ in terms of the Fried parameter $r_0$,
\begin{equation}
\int_0^z h(z')\ {\rm d}z' = 0.592 \left( \frac{\omega_0}{r_0} \right)^{5/3} ,
\label{ihaaz}
\end{equation}
where $r_0 = 0.185(\lambda^2/C_n^2/z)^{3/5}$. Thus all the dimension parameters are combined into $\omega_0/r_0$.

\begin{table}
\caption{\label{indkon}The numerical values for the constants $A_q$ and $B_q$ that appear in $L_{mnrs}$ for $q=1,2,3$.}
\begin{tabular}{c|lll}
\hline
\hline
 & $q=1$ & $q=2$ & $q=3$ \\
\hline
$A_q$ & 1.570 & 2.206 & 2.807 \\
$B_q$ & 0.03030 & 0.004787 & 0.001754 \\
\hline
\hline
\end{tabular}
\end{table}

\begin{figure}[ht]
\centerline{\scalebox{1}{\includegraphics{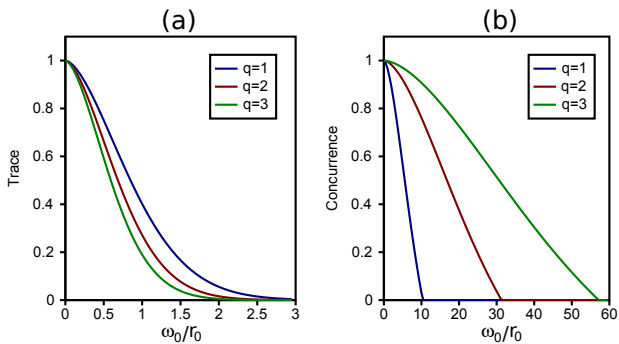}}}
\caption{Plots of (a) the trace of the density matrix and (b) the concurrence for a biphoton, initially in the singlet Bell-state, in terms of two OAM states with $l=\pm q$, for $q=1,2,3$, as a function of $\omega_0/r_0$.}
\label{conc}
\end{figure}

Assuming that the initial state of the density matrix is the singlet Bell-state in the OAM basis, one obtains the following solution of the density matrix
\begin{equation}
\rho_{mnpq} = \frac{T}{4} \left[ \begin{array}{cccc}
1-R^2 & 0 & 0 & 0 \\
0 & 1+R^2 & -2R & 0 \\
0 & -2R & 1+R^2 & 0 \\
0 & 0 & 0 & 1-R^2 \\
\end{array} \right] ,
\label{densmat}
\end{equation}
where $mp$ ($nq$) denote the row (column) indices, and where
\begin{eqnarray}
T & = & \exp \left[-(A_q-B_q) \int_0^z h(z')\ {\rm d}z' \right] \label{tee} \\
R & = & \exp \left[-B_q \int_0^z h(z')\ {\rm d}z' \right] . \label{er}
\end{eqnarray}
The eigenvalues of the density matrix are $T(1+R)^2/4$, $T(1-R)^2/4$, $T(1-R^2)/4$ and $T(1-R^2)/4$, which are all positive. The trace of the density matrix is given by $T$, which is a decaying function, since $A_q>B_q$, as shown in Table~\ref{indkon}. We plot the trace $T$ as a function of $\omega_0/r_0$ in Fig.~\ref{conc}(a) for $q=1,2,3$.

Using the normalized density matrix [by setting $T=1$ in Eq.~(\ref{densmat})], we compute the concurrence of formation \cite{wootters1,wootters2} and obtain ${\cal C} = (2R+R^2-1)/2$, which is plotted in Fig.~\ref{conc}(b) for $q=1,2,3$ as a function of $\omega_0/r_0$.

From the curves for the trace in Fig.~\ref{conc}(a) one can see that modes with higher OAM are scattered more rapidly into other modes than those with lower OAM. On the other hand, from Fig.~\ref{conc}(b) we see that modes with higher OAM retain their entanglement for longer distances than those with lower OAM. These conclusions agree qualitatively with previous work \cite{qturb1}, however, while the scattering into other modes occurs at a scale where $r_0\approx\omega_0$, similar to what was found before \cite{qturb1}, the entanglement lasts for at least an order of magnitude longer, which is quantitatively different from what was found before \cite{qturb1}. Here the slowness of the decay in the concurrence is a result of the smallness of the values of the $B_q$'s given in Table~\ref{indkon}. From these results it appears that the effect of scattering and the implied loss of photons in the desired OAM modes may turn out to be a more significant challenge for free-space quantum communication than the decoherence of OAM entanglement.

In conclusion, we derived an infinitesimal propagator equation, reminiscent of a Master equation, that models the spatial evolution of a density operator for a OAM entangled multiphoton state, propagating through a turbulent atmosphere. The approach assumes monochromatic light in the paraxial limit, but does not require any simplification to the turbulence model. As a result this formulation is capable of analyzing the effect of realistic atmospheric turbulence on the propagation of any quantum photon state in the OAM basis. The resulting infinite set of first order differential equation was truncated to consider an example for comparison with previous work. Although the result was found to be qualitatively similar, significant quantitative differences exist.

The author is grateful for the discussions he had with Andrew Forbes, Thomas Konrad, Francesco Petruccione and Hermann Uys on this topic. This work was done with the financial support of an SRP Type A grant from the CSIR.


\end{document}